\documentclass[showpacs,twocolumn,prb,eqsecnum,amsmath,amssymb]{revtex4}
\usepackage{graphicx}
\usepackage{dcolumn}
\usepackage{bm}
\begin{document}
\title{
Origin and roles of a strong electron-phonon interaction\\
in cuprate oxide superconductors
}
\author{Fusayoshi J. Ohkawa}
\affiliation{Department of Physics, Faculty of
Science,  Hokkaido University, Sapporo 060-0810, Japan}
\email{fohkawa@phys.sci.hokudai.ac.jp}
\received{21 July 2006; revised manuscript received 1 January 2007; accepted for publication in PRB}
\begin{abstract} 
 
A strong electron-phonon interaction arises from
the modulation of the superexchange interaction by lattice vibrations.
It is responsible for the softening of the  half-breathing modes
around $(\pm \pi/a,0)$ and $(0,\pm \pi/a)$ in the two-dimensional
Brillouin zone, with $a$ being the lattice constant of CuO$_2$ planes,
as is studied in Phys.\! Rev.\! B {\bf 70}, 184514 (2004).
Provided that antiferromagnetic spin fluctuations are
developed around ${\bf Q}=\left(\pm 3\pi/4a, \pm\pi/a\right)$ and
$\left(\pm\pi/a, \pm 3\pi/4a \right)$,
the electron-phonon interaction
can also cause the softening of Cu-O bond stretching modes 
around $2{\bf Q}$, or around
$(\pm \pi/2a,0)$ and $(0,\pm \pi/2a)$.
The softening  
around $2{\bf Q}$ is accompanied  by the development of  
charge fluctuations corresponding to the so called 
$4a$-period stripe or 
$4a\hspace{-1pt} \times\hspace{-1pt} 4a$-period checker-board state.
However, an observation that
the $4a$-period modulating part or the $2{\bf Q}$
part of the density of states is 
almost symmetric with respect to the chemical potential
contradicts a scenario that the stabilization of 
a single-$2{\bf Q}$ or double-$2{\bf Q}$ charge density wave 
following the complete softening of the $2{\bf Q}$ bond stretching modes
is responsible for the ordered stripe or checker-board state.
It is proposed that the stripe or checker-board state is simply 
a single-{\bf Q} or double-{\bf Q} spin density wave, whose
second-harmonic effects can explain
the observed almost symmetric $2{\bf Q}$ part of the density of states.
The strong electron-phonon interaction can play no or only a minor role
in the occurrence of $d\gamma$-wave superconductivity
in cuprate oxides. 
\end{abstract}
\maketitle

\section{Introduction}

It is one of the most interesting and 
important issues in condensed-matter physics
to elucidate the mechanism of high critical temperature (high-$T_c$)
superconductivity occurring in cuprate oxides. \cite{bednorz} 
The oxides are highly anisotropic quasi-two-dimensional oxides, whose
main compositions are CuO$_2$ planes.
High-$T_c$ superconductivity occurs on the CuO$_2$ planes.
There are  pieces of evidence that the electron-phonon interaction is strong
on the CuO$_2$ planes:
the softening of the half-breathing modes
around $\left(\pm \pi/a, 0\right)$ and $\left(0, \pm \pi/a\right)$
in the two-dimensional Brillouin zone (2DBZ),
\cite{McQ1,Pint1,McQ2,Pint2,Braden}
with $a$ being the lattice constant of the CuO$_2$ planes,
the softening of Cu-O bond stretching modes
around $(\pm \pi/2a,0)$ and $(0,\pm \pi/2a)$ in 2DBZ, 
\cite{pintschovius,reznik} 
kinks in the dispersion relation of quasiparticles, 
\cite{johnson,tsato} and so on.
It may be argued therefore that the electron-phonon interaction must
play a major role in the occurrence of high-$T_c$ superconductivity.
On the other hand, observed  isotope shifts of 
$T_c$ are small, \cite{isotope} which implies that
the strong electron-phonon interaction can play only a minor role
in high-$T_c$ superconductivity itself. 
The origin and roles of the strong electron-phonon interaction
should be clarified in order that the issue 
of high-$T_c$ superconductivity might be solved.

Parent cuprate oxides with no doping are Mott insulators.
When holes or electrons are doped into the the Mott insulators,
high-$T_c$ superconductivity appears.
Cuprate oxide superconductors lie 
in the vicinity of the Mott metal-insulator transition or crossover.
It may be argued therefore that
strong electron correlations must play a crucial role
not only in the occurrence of high-$T_c$ superconductivity
but also in the origin and roles of the strong electron-phonon interaction.

The  Hubbard model is one of the simplest effective Hamiltonians
for strongly correlated electron liquids.
In Hubbard's approximation, \cite{Hubbard1,Hubbard2}
a band splits into two subbands
when the on-site repulsion $U$ is so large 
that $U\agt W$, with $W$ being the bandwidth of unrenormalized electrons. 
The subbands are called the upper Hubbard band (UHB)
and the lower Hubbard band (LHB), and
a gap between UHB and LHB is called the Hubbard gap.
In Gutzwiller's approximation, \cite{Gutzwiller1,Gutzwiller2,Gutzwiller3}
a narrow quasiparticle band appears around the chemical potential.
The band and quasiparticles are called the Gutzwiller band 
and quasiparticles, respectively. 
It is plausible to speculate that 
the density of states has in fact a three-peak structure,
with the Gutzwiller band between UHB and LHB. 
Both of the approximations 
are single-site approximations (SSA).
Another SSA theory  confirms the speculation, \cite{OhkawaSlave}
showing that the Gutzwiller band appears
at the top of LHB when the electron density per unit cell 
is less than one. 
The nature of  the ground state of the Hubbard model
depends on the nature of the Gutzwiller quasiparticles.

The SSA that considers all the single-site terms
is reduce to determining and solving self-consistently
the Anderson model, \cite{Mapping-1,Mapping-2,Mapping-3}
which  is one of the simplest effective Hamiltonian for the Kondo problem.
Hence, the three-peak structure corresponds to the Kondo peak
between two subpeaks in the Anderson model, or in the Kondo problem.
The $s$-$d$ model is also one of the simplest effective Hamiltonian for 
the Kondo problem.
According to Yosida's perturbation theory \cite{yosida} and Wilson's
renormalization-group theory,  \cite{wilsonKG}  
the ground state of the $s$-$d$ model  is a singlet 
or a normal Fermi liquid
provided that the Fermi surface of conduction electrons is present.
Since  the $s$-$d$ model is derived from the Anderson model,
the ground state of the Anderson model  is also a normal Fermi liquid. 
It is certain therefore that under the SSA 
the ground state of the Hubbard model is 
a normal Fermi liquid or a metal.\cite{ohkawaM-I}
Even if the Hubbard gap opens,  
the Fermi surface of the Gutzwiller quasiparticles is present.

The SSA can also be formulated as
the dynamical mean-field theory \cite{georges} 
and the dynamical coherent potential approximation. \cite{kakehashi}
In the SSA,
local fluctuations are rigorously considered but Weiss mean fields,
which are responsible for the appearance of the corresponding order
parameter, are ignored. Hence, 
the SSA is rigorous for infinite dimensions within the
Hilbert subspace with no order parameter. \cite{Metzner}  
In Kondo-lattice theory, \cite{Mapping-1,Mapping-2,Mapping-3}
an {\it unperturbed} state is the normal Fermi liquid, which is
constructed in the non-perturbative SSA theory, and 
not only effects of intersite fluctuations but also ordering due to 
Weiss mean fields such as magnetism, superconductivity and so on
are perturbatively considered. 
Kondo-lattice theory can also be formulated as $1/d$ expansion theory,
with $d$ being the spatial dimensionality.

The $d$-$p$ model, 
where Cu $3d$ and O $2p$ orbits are explicitly considered,  is
one of the simplest effective Hamiltonians for
cuprate oxide superconductors.
Since the on-site repulsion $U$ plays a crucial role 
in the $d$-$p$ model as it does in the Hubbard model,
it is straightforward to extend the analysis for the Hubbard model
to the $d$-$p$ model.
Observed quasiparticle states, which are often called
mid-gap states,  are simply the Gutzwiller quasiparticle states, which
can also be renormalized by intersite fluctuations.
When observed specific heat coefficients  
as large as 
%
$\gamma \simeq 14~\mbox{mJ/K$^2$mol}$
%
are used, \cite{gamma1,gamma2} 
the Fermi-liquid relation  gives \cite{Luttinger1,Luttinger2}
\begin{equation}\label{EqGamma2}
W^*=\mbox{0.3-0.4~eV},
\end{equation}
for the effective bandwidth of the Gutzwiller quasiparticles 
in optimal-doped cuprate oxide superconductors, where
$T_c$ is the highest as a function of doping concentrations.
According to field theory, the superexchange interaction
arises from the virtual exchange of a pair excitation
of electrons across the Hubbard gap. \cite{OhSupJ1}
Since the Gutzwiller quasiparticles, which are responsible for
metallic properties, 
plays no significant role in the virtual exchange process, 
the superexchange interaction is relevant even in a metallic phase,
provided that the Hubbard gap opens. 
Cooper-pairs can also be bound 
by a magnetic  exchange interaction. \cite{hirsch} 
Since the superexchange interaction constant is as strong as  \cite{SuperJ}
%
$J=- \mbox{(0.10-0.15)~eV}$  
%
between nearest-neighbor Cu ions on a CuO$_2$ plane,
observed high $T_c$ can be easily reproduced.
In actual, it has already been proposed that
the condensation of $d\gamma$-wave Cooper pairs between
the Gutzwiller quasiparticles  
due to the superexchange interaction 
is responsible for high-$T_c$ superconductivity.\cite{highTc1,highTc2}
Since the superexchange interaction is strong 
only between nearest-neighbor Cu ions,
it is definite that theoretical $T_c$ of the $d\gamma$ wave
is much higher than those of other waves. In fact,
high-$T_c$ superconductivity occurs in an intermediate-coupling regime
$|J|/W^* = 0.3\mbox{-}0.5$
for superconductivity, which is realized 
in the strong-coupling regime for electron correlations
defined by $U/W \agt 1$. 

Since charge fluctuations are suppressed by strong electron correlations,
the conventional electron-phonon
interaction arising from charge-channel interactions must be
weak in cuprate oxide superconductors. On the other hand,
an electron-phonon interaction arising  from spin-channel interactions 
can be strong. For example, an electron-phonon
interaction arising from the modulation of a magnetic exchange
interaction by phonons plays a significant role 
in the spin-Peierls effect. It has been shown in a previous paper
\cite{novel-el-ph}
that an electron-phonon interaction arising from the modulation of 
the superexchange interaction by phonons is strong in cuprate oxide
superconductors. The electron-phonon interaction
can explain the softening of the half-breathing modes
around $\left(\pm \pi/a, 0\right)$
and $\left(0, \pm \pi/a\right)$ in 2DBZ.
It has been predicted that the softening must be small around 
$\left(\pm \pi/a, \pm \pi/a\right)$  in 2DBZ.
An attractive mutual interaction 
due to such an electron-phonon interaction
is strong between quasi-particles on next-nearest-neighbor Cu ions, but
is very weak between those on nearest-neighbor Cu ions.
Therefore, the mutual interaction can play no significant role 
in the binding of $d\gamma$-wave Cooper pairs. 
Observed small isotope shifts of $T_c$ can never contradict
the presence of the strong electron-phonon interaction.

The so called $4a$-period stripes and $4a\!\times\! 4a$-period
checker boards are observed in
under-doped cuprate oxide superconductors,
\cite{howald,Vershinin,Hanaguri,Momono,McElroy,kiverson}
whose doping concentrations
are smaller than those of optimal-doped ones.
The wave numbers of Cu-O bond stretching modes, 
$(\pm \pi/2a,0)$ and $(0,\pm \pi/2a)$  in 2DBZ,
correspond to the period $4a$ of  stripes and checker boards.
The softening of the stretching modes is accompanied by 
the development of $4a$-period or $4a\!\times\! 4a$-period
fluctuations in charge channels, which are simply 
stripe or checker-board fluctuations.
It may be argued therefore that a charge density wave (CDW)
following the complete softening
of the bond stretching modes is responsible for 
ordered stripes and checker boards.

One of the purposes of this paper is to show that
the strong electron-phonon interaction can also explain the softening 
of Cu-O bond stretching modes in cuprate oxide superconductors.
The other purpose is to examine critically
the relevance of the CDW scenario,  
whether or not the CDW
is actually responsible for ordered stripes and checker boards.
This paper is organized as follows:
Preliminary discussions are presented in Sec.~\ref{SecPrelim};
the derivation of the electron-phonon interaction 
is reviewed in Sec.~\ref{EP-interaction} and
Kondo-lattice theory is reviewed in Sec.~\ref{KL-theory}. 
The softening of the bond stretching modes  around 
$(\pm \pi/2a,0)$ and $(0,\pm \pi/2a)$ in 2DBZ is studied 
in Sec.~\ref{SecSoftening}. 
The relevance of the CDW scenario 
for stripes and checker boards
is critically examined in Sec.~\ref{stripes}. 
An argument on the mechanism of high-$T_c$ superconductivity
is given in Sec.~\ref{SecDisuss}.
Conclusion is presented in Sec.~\ref{SecConclusion}.

\section{Preliminaries}\label{SecPrelim}
\subsection{Electron-phonon interaction}
\label{EP-interaction}

In cuprate oxide superconductors,
the superexchange interaction arises from the virtual exchange of a pair 
excitation of $3d$ electrons between UHB and LHB that
are strongly hybridized subbands between Cu $3d$ and O $2p$ orbits.
\cite{OhSupJ1}
When the broadening or finite bandwidths of UHB and LHB are ignored,
the exchange interaction constant  
between nearest-neighbor Cu ions on a CuO$_2$ plane is given by
\begin{equation}\label{EqSuperJ-1}
J =-  \frac{4V^4}{(\epsilon_d-\epsilon_p+U)^2}
\left(\frac{1}{\epsilon_d-\epsilon_p+U}+\frac{1}{U} \right) ,
\end{equation}
with $V$ being
the hybridization matrix between nearest-neighbor O $2p$ and
Cu $3d$ orbits, $\epsilon_d$ and $\epsilon_p$ the depths of 
Cu $3d$ and O $2p$ levels, and $U$
the on-site repulsion between Cu $3d$ electrons.

Doped holes reside mainly at O ions.
The preferential doping suggests that O $2p$ levels are shallower than
Cu $3d$ levels or that
parent cuprate oxides with no hole  doping must be
charge-transfer insulators rather than Mott insulators;
charge-transfer insulators and Mott insulators are characterized by
$\epsilon_p > \epsilon_d$ and $\epsilon_p < \epsilon_d$, respectively.
Since the hybridization between Cu $3d$ and O $2p$ orbits
must be strong, it may also be argued that Cu $3d$ levels are
much deeper than O $2p$ levels, that is, 
$\epsilon_p \gg \epsilon_d$ rather than $\epsilon_p > \epsilon_d$,
to explain the observed preferential doping.
However, the suggested level scheme 
of $\epsilon_p > \epsilon_d$ or 
$\epsilon_p \gg \epsilon_d$ disagrees with the prediction of 
Mott insulators,
$\epsilon_{d}-\epsilon_{p} \simeq 1\mbox{~eV}$,
by band calculations. \cite{band1,band2,band3}
The preferential doping 
does not necessarily mean that the parent cuprate oxides are
charge-transfer insulators, but it
simply means that the local charge susceptibility of $3d$ electrons is much
smaller than  that of $2p$ electrons, which implies that the effective on-site
repulsion $U$ between $3d$ electrons is very strong. 
It is assumed in this paper that
$V\simeq1.6$~eV and $\epsilon_{d}-\epsilon_{p}\simeq$1~eV, as is
predicted by band calculations. \cite{band1,band2,band3}
Since the on-site $U$ should be so large that
the Hubbard gap might open, it is assumed that
$U\simeq5$~eV. Then, Eq.~(\ref{EqSuperJ-1}) gives
$J \simeq - 0.27~\mbox{eV}$. 
This is about twice as large as the experimental 
$J = - (0.10\mbox{--}0.15)~\mbox{eV}$. \cite{SuperJ}
This discrepancy is resolved when nonzero bandwidths of UHB and LHB
are considered. \cite{OhSupJ1}

Displacements of the $i$th Cu ion and the $[ij]$th O ion,
which lies between the nearest-neighbor $i$th and $j$th Cu ions, are given by
\begin{equation}\label{EqDispCu}
{\bf u}_i = 
 \sum_{\lambda{\bf q}}
 \frac{\hbar v_{d,\lambda{\bf q}} } 
{\sqrt{ 2NM_d \omega_{\lambda{\bf q}}} } 
e^{i{\bf q}\cdot{\bf R}_i}
{\bm \epsilon}_{\lambda{\bf q}} \left(
b_{\lambda-{\bf q}}^\dag \!+\! b_{\lambda{\bf q}} \right), 
\end{equation}
and
\begin{equation} \label{EqDispO}
{\bf u}_{[ij]} =
\sum_{\lambda{\bf q}}
\frac{\hbar v_{p,\lambda{\bf q}}}
{\sqrt{2N M_p \omega_{\lambda{\bf q}}} } 
 e^{i{\bf q}\cdot {\bf R}_{[ij]} }
{\bm \epsilon}_{\lambda{\bf q}} \! \left(
b_{\lambda-{\bf q}}^\dag \!\!+\! b_{\lambda{\bf q}} \right), 
\end{equation}
with  ${\bf R}_i$ and 
${\bf R}_{[ij]} = ({\bf R}_i + {\bf R}_j)/2$
being positions of the $i$th Cu  and $[ij]$th O ions, 
$M_d$ and $M_p$ masses of Cu and O ions, 
$b_{\lambda{\bf q}}^\dag$ and $b_{\lambda{\bf q}}$
creation and annihilation operators of a phonon with 
a polarization $\lambda$ and a wave vector ${\bf q}$, 
$\omega_{\lambda{\bf q}}$ a phonon energy,  
%
${\bm \epsilon}_{\lambda{\bf q}}
=(\epsilon_{\lambda {\bf q},x},
\epsilon_{\lambda {\bf q},y},\epsilon_{\lambda {\bf q},z})$
a polarization vector, and $N$ the number of unit cells. 
Here, only longitudinal phonons are considered so that 
it is assumed that
$\epsilon_{\lambda{\bf q}} = (q_x,q_y,q_z)/|{\bf q}| $
for ${\bf q}$ within the first Brillouin zone.
The ${\bf q}$ dependence of
$v_{d,\lambda{\bf q}}$ and $v_{p,\lambda{\bf q}}$ is crucial.
For example, 
$v_{d,\lambda{\bf q}}=0$ and 
$v_{p,\lambda{\bf q}} = O(1)$   
for modes that bring no change in adjacent Cu-Cu distances.

Two types of electron-phonon interactions
arise from the modulations of the superexchange interaction $J$ 
by the vibrations of O and Cu ions. When they are considered,
it is convenient to define a {\it dual-spin} operator.
First, a {\it  single-spin} operator is defined by
\begin{equation}
{\bf S}({\bf q})= \frac1{\sqrt{N}} \sum_{\bf k\alpha\beta} 
\frac1{2}{\bm \sigma}^{\alpha\beta} d_{({\bf k} +\frac{1}{2}{\bf q})
\alpha}^\dag   d_{({\bf k} -\frac{1}{2}{\bf q}) \beta}, 
\end{equation}
with  
$ {\bm \sigma} = \left(\sigma_x , \sigma_y ,\sigma_z \right)$
being the Pauli matrixes and 
$d_{{\bf k}\sigma}^\dag$ and $d_{{\bf k}\sigma}$
being creation and annihilation operators, respectively, of $3d$ electrons 
with wave number ${\bf k}$. Then, 
the dual-spin operator is defined by
\begin{equation}\label{EqTwoSpin}
{\cal P}_\Gamma({\bf q}) = \frac1{2} 
\sum_{{\bf q}^\prime}\eta_{\Gamma}({\bf q}^\prime) \! \left[
 {\bf S}\left({\bf q}^\prime \!\!+\! \mbox{$\frac{1}{2}$}{\bf q}\right)
\!\cdot {\bf S}\left(-{\bf q}^\prime \!\!+\! \mbox{$\frac{1}{2}$}{\bf q}
\right)\right] ,
\end{equation}
with
\begin{equation}
\eta_{s}({\bf q}) = \cos(q_xa) + \cos(q_ya),
\end{equation}
and 
\begin{equation}
\eta_{d}({\bf q}) = \cos(q_xa) - \cos(q_ya) .
\end{equation}
It is assumed in this paper 
that the $x$ and $y$ axes are within CuO$_2$ planes and
the $z$ axis is perpendicular to CuO$_2$ planes.
The electron-phonon interactions are simply given by
\begin{eqnarray}\label{EqElPhP}
{\cal H}_p &=&
i C_p \sum_{\bf q} 
\frac{\hbar v_{p,\lambda{\bf q}}}
{\sqrt{2 N M_p \omega_{\lambda{\bf q}}}} 
\left(b_{\lambda-{\bf q}}^\dag + b_{\lambda{\bf q}} \right)
\nonumber \\ &&  \quad \times
\bar{\eta}_{s}({\bf q}) \sum_{\Gamma=s,d} 
\eta_{\Gamma}\left(\mbox{$\frac{1}{2}{\bf q}$}\right) 
{\cal P}_\Gamma({\bf q}) , 
\end{eqnarray}
and
\begin{eqnarray} \label{EqElPhD}
{\cal H}_d &=&
i C_d \sum_{\bf q} 
\frac{\hbar v_{d,\lambda{\bf q}}}
{\sqrt{2 N M_d \omega_{\lambda{\bf q}}}} 
\left(b_{\lambda-{\bf q}}^\dag + b_{\lambda{\bf q}} \right)
\nonumber \\ &&  \quad \times
\sum_{\Gamma=s,d} 
\bar{\eta}_{\Gamma}({\bf q})
{\cal P}_\Gamma({\bf q}) , 
\end{eqnarray}
with $C_p$ and $C_d$ being real constants, 
which are given in the previous paper,
\cite{novel-el-ph} and
%
%
%
%
\begin{equation}
\bar{\eta}_{s}({\bf q}) =
2\left[ e_{\lambda x} \sin\left(\frac{q_x a}{2}\right) 
+  e_{\lambda y} \sin\left(\frac{q_y a}{2}\right)\right] ,
\end{equation}
and
\begin{equation}
\bar{\eta}_{d}({\bf q}) =
2\left[ e_{\lambda x} \sin\left(\frac{q_x a}{2}\right) 
-  e_{\lambda y} \sin\left(\frac{q_y a}{2}\right)\right] .
\end{equation}
%

\subsection{Kondo-lattice theory}
\label{KL-theory}

One of the simplest effective Hamiltonians for the electron part
of cuprate oxide superconductors is the $d$-$p$ model on a  square lattice.
Since the anisotropy is large,
it is convenient to consider phenomenologically
quasi-two-dimensional features.
The $d$-$p$ model  is approximately mapped to  the $t$-$J$ model or
the $t$-$J$-infinite-$U$ model: \cite{ZhangRice} 
\begin{eqnarray}\label{EqtJ}
{\cal H}_{t\mbox{-}J} &=& 
\sum_{ij\sigma} t_{ij} d_{i\sigma}^\dag d_{j\sigma} 
-  \frac1{2}J 
\sum_{\left<ij\right>} ({\bf S}_i \cdot {\bf S}_j)  
\nonumber \\ && \quad 
+  U_{\infty} \sum_{i}  
d_{i\uparrow}^\dag d_{i\uparrow}
d_{i\downarrow}^\dag d_{i\downarrow} ,
\end{eqnarray}
with  the summation over $\left<ij\right>$ indicating that
the summation should be made over nearest neighbors and
\begin{equation}\label{EqSi}
{\bf S}_i  = \sum_{\alpha\beta}   \frac1{2}
{\bm \sigma}^{\alpha\beta}  d_{i\alpha}^\dagger d_{i\beta}.
\end{equation}
The carrier density per unit cell is defined by 
\begin{equation}
n= \frac1{N} \sum_{i}
\bigl<d_{i\sigma}^\dag d_{i\sigma}\bigr>. 
\end{equation}
It should be noted that
the infinitely large on-site repulsion $U_\infty$ is introduced
to exclude double occupancy so that $n$ can never be larger than
unity, or  $0\le n \le 1$.
The electron and hole pictures should be taken for the so called 
{\it hole}-doped $(n<1)$ and {\it electron}-doped $(n>1)$ cuprate oxide superconductors,
respectively, so that $n$ is the electron density
for {\it hole}-doped ones and is the hole density 
for {\it electron}-doped ones.
The doping concentration is defined by $\delta= 1-n$, and
the optimal concentration, 
where superconducting $T_c$ is the highest as
a function of $\delta$,  is $\delta \simeq0.15$.
Then, $\delta\alt 0.15$ and 
$\delta \agt 0.15$ are called under-doped and over-doped
concentrations, respectively.
When transfer integrals between nearest and next-nearest neighbors,
which are denoted by $t$ and $t^\prime$, are only considered,
the dispersion relation of electrons or holes is given by
\begin{eqnarray}
%
E({\bf k}) &=&
2 t \left[\cos(k_xa) + \cos(k_ya)\right]
\nonumber \\ && \quad 
+ 4 t^\prime \cos(k_xa)\cos(k_ya).
\end{eqnarray}
According to band calculations,\cite{band1,band2,band3} 
it follows that
$t=-(0.3\mbox{-}0.5)$~eV and $t^\prime \simeq - 0.3 t$
for electrons in hole-doped cuprate oxide superconductors
and $t=-(0.3\mbox{-}0.5)$~eV and $t^\prime \simeq + 0.3 t$
for holes in electron-doped ones.

Every physical quantity is divided into single-site and multi-site terms.
Calculating the single-site term is  reduced to
determining and solving selfconsistently the Anderson model,
as is discussed in the Introduction.
When it is assumed that there is no order parameter, for example,
the self-energy of electrons 
is divided into single-site and multi-site self-energies:
\begin{equation}
\Sigma_{\sigma}(i\varepsilon_n, {\bf k})
= \tilde{\Sigma}_{\sigma}(i\varepsilon_n)
+ \Delta \Sigma_{\sigma}(i\varepsilon_n, {\bf k}).
\end{equation}
The single-site self-energy  $\tilde{\Sigma}_{\sigma}(i\varepsilon_n)$
is given by that of the Anderson model.
It is expanded as
\begin{eqnarray}\label{EqExpandSelf}
\tilde{\Sigma}_\sigma(\varepsilon  +  i0)  &=&
\tilde{\Sigma}_0
+ \bigl(1  -  \tilde{\phi}_\gamma \bigr)\varepsilon 
\nonumber \\ && 
+ \bigl(1  - \tilde{\phi}_s \bigr)  \frac1{2}\sigma g \mu_BH 
 + O(\varepsilon^2), \qquad 
\end{eqnarray}
at $T=0$~K
in the presence of an infinitesimally small Zeeman energy $g\mu_BH$,
with
$g$ the $g$ factor and $\mu_B$ the Bohr magneton.
The expansion coefficients $\tilde{\Sigma}_0$, $\tilde{\phi}_\gamma$, 
and $\tilde{\phi}_s$  are all real; 
$\tilde{\phi}_s \simeq 2 \tilde{\phi}_\gamma \gg 1$
for $n\simeq 1$.
When Eq.~(\ref{EqExpandSelf}) is used and
the multi-site self-energy is ignored,
the dispersion relation of the Gutzwiller quasiparticles is given by
\begin{equation}
\xi_\sigma({\bf k}) = \frac1{\tilde{\phi}_\gamma}
\left[\tilde{\Sigma}_0 +E({\bf k}) -\mu \right] 
-\frac1{2}\sigma \tilde{W}_s g\mu_B H,
\end{equation}
with
\begin{equation}
\tilde{W}_s = \tilde{\phi}_s/\tilde{\phi}_\gamma ,
\end{equation}
being the so called Wilson ratio for the Kondo problem.

The irreducible polarization function in spin channels
is also divided into single-site and multi-site polarization functions:
\begin{equation}
\pi_s(i\omega_l,{\bf q}) =
\tilde{\pi}_s(i\omega_l) +\Delta\pi_s(i\omega_l,{\bf q}) .
\end{equation}
The single-site polarization function $\tilde{\pi}_s(i\omega_l)$
is given by that of the Anderson model.
The spin susceptibilities of the Anderson and 
$t$-$J$ models are given, respectively, by
\begin{equation}\label{EqChi-tJ}
\tilde{\chi}_s(i\omega_l) =
\frac{2\tilde{\pi}_s(i\omega_l)}{
1 - U_\infty \tilde{\pi}_s(i\omega_l)},
\end{equation}
and
\begin{equation}\label{EqChi-A}
\chi_s(i\omega_l,{\bf q}) =
\frac{2\pi_s(i\omega_l,{\bf q})}{
1 - \left[\frac1{4}J({\bf q}) \!+\! U_\infty 
\right]\pi_s(i\omega_l,{\bf q})},
\end{equation}
with
\begin{equation}\label{EqSuperJ-2}
J({\bf q}) = 2 J  
\left[\cos(q_xa) + \cos(q_ya) 
\right]  .
\end{equation}
In Eqs.~(\ref{EqChi-tJ}) and (\ref{EqChi-A}),
the conventional factor
$\frac1{4}g^2\mu_B^2$ is not included.
A physical picture for Kondo lattices is that
local spin fluctuations at different sites interact with each other
by an intersite exchange interaction.
In Kondo-lattice theory,  therefore, 
an intersite exchange interaction 
$I_s(i\omega_l,{\bf q})$ is defined by
\begin{equation}\label{EqSus}
\chi_s(i\omega_l,{\bf q}) =
\frac{\tilde{\chi}_s(i\omega_l)}{  
1 - \frac1{4}I_s(i\omega_l,{\bf q}) \tilde{\chi}_s(i\omega_l)} .
\end{equation}
It follows that
\begin{equation}\label{Eq-I}
I_s(i\omega_l,{\bf q}) = J({\bf q})
+ 2 U_\infty^2 \Delta\pi_s(i\omega_l,{\bf q}) .
\end{equation}
The derivation of Eq.~(\ref{Eq-I}) from Eqs.~(\ref{EqChi-tJ}) and (\ref{EqChi-A})
is rigorous because ignored terms, which are  
$O[1/U_\infty \tilde{\chi}_s(i\omega_l)]$, vanish
for infinitely large $U_\infty$.
The term of $2 U_\infty^2 \Delta\pi_s(i\omega_l,{\bf q})$
is composed of two terms: \cite{FJO-disorder}
\begin{equation}
2 U_\infty^2 \Delta\pi_s(i\omega_l,{\bf q}) =
J_Q(i\omega_l,{\bf q}) - 4 \Lambda (i\omega_l,{\bf q}).
\end{equation}
The first term $J_Q(i\omega_l,{\bf q})$ 
is an exchange interaction arising from the virtual exchange
of a pair excitation of the Gutzwiller quasiparticles.
According to the Ward relation, \cite{ward}
the static component of the single-site
irreducible vertex function in spin channels
is given by 
\begin{equation}\label{EqWard1}
\tilde{\lambda}_s = \tilde{\phi}_s[1 -U_\infty \tilde{\pi}_s(0)] .
\end{equation}
Then, it follows that
\begin{equation}\label{EqWard2}
U_\infty\tilde{\lambda}_s 
= 2\tilde{\phi}_s/\tilde{\chi}_s(0).
\end{equation}
When the vertex correction $\tilde{\lambda}_s $ 
given by Eq.~(\ref{EqWard2}) is 
used, it follows that
\begin{equation}\label{EqJQ}
J_Q(i\omega_l,{\bf q}) =
\frac{4\tilde{W}_s^2}{\tilde{\chi}_s^2(0)} 
\left[
P(i\omega_l,{\bf q}) - \frac1{N}\sum_{\bf q} P(i\omega_l,{\bf q})
\right],
\end{equation}
with
\begin{equation}\label{EqP}
P(i\omega_l,{\bf q}) =
\frac1{N}\sum_{{\bf k}\sigma}
\frac{f[\xi_\sigma({\bf k})]-f[\xi_\sigma({\bf k}+{\bf q}) ]}
{\xi_\sigma({\bf k}+{\bf q}) - \xi_\sigma({\bf k})
+i\omega_l} ,
\end{equation}
with
\begin{equation}
f(\varepsilon) = \frac1{e^{\varepsilon/k_BT} +1} .
\end{equation}
In Eq.~(\ref{EqJQ}), the single-site term is subtracted because
it is considered in the SSA.
The strength of the exchange interaction
is proportional to the width of the Gutzwiller band.
Since the chemical potential lies around the center of the Gutzwiller band
or the nesting of the Fermi surface is sharp,
the exchange interaction is antiferromagnetic (AF) 
in cuprate oxide superconductors.
The second term 
$ - 4\Lambda (i\omega_l,{\bf q})$ corresponds to the mode-mode
coupling term in the self-consistent renormalization  theory
of spin fluctuations, \cite{moriya} which
is relevant in the weak-coupling regime for electron correlations
defined by $U/W \alt 1$.

When Eq.~(\ref{EqWard2}) is used, 
the mutual interaction mediated by intersite spin fluctuations,
which works between the Gutzwiller quasiparticles, is given by
\begin{equation}\label{EqSF}
(U_\infty \tilde{\lambda}_s)^2
[\chi_s(i\omega_l,{\bf q}) - \tilde{\chi}_s(i\omega_l)]=
\tilde{\phi}_s^2 I_s^*(i\omega_l,{\bf q}), 
\end{equation}
with
\begin{equation}
I_s^*(i\omega_l,{\bf q})=\frac{I_s(i\omega_l,{\bf q})}
{1 - \frac1{4} I_s(i\omega_l,{\bf q}) \tilde{\chi}_s(i\omega_l)}.
\end{equation}
In Eq.~(\ref{EqSF}), 
the single-site term is subtracted because it is considered 
in the SSA. The exchange interaction $I_s(i\omega_l,{\bf q})$
is enhanced into $I_s^*(i\omega_l,{\bf q})$
by spin fluctuations.
The mutual interaction mediated by spin fluctuations
is essentially the same as that due to the exchange interaction
$I_s(i\omega_l,{\bf q})$ or $I_s^*(i\omega_l,{\bf q})$.

In Kondo-lattice theory, 
a {\it unperturbed} state is constructed in
the non-perturbative SSA theory and
multi-site or intersite effects are perturbatively considered
in terms of $I_s(i\omega_l,{\bf q})$ or $I_s^*(i\omega_l,{\bf q})$.

\section{Softening of phonons
due to antiferromagnetic spin fluctuations}
\label{SecSoftening}

An  effective Hamiltonian to be eventually examined in this paper is
\begin{equation}\label{EqHamiltonian}
{\cal H}= {\cal H}_{t\mbox{-}J} + {\cal H}_{\rm ph}
+ {\cal H}_p + {\cal H}_d , 
\end{equation}
with 
%
\begin{equation} 
{\cal H}_{\rm ph}=\sum_{\lambda{\bf q}}
\omega_{\lambda{\bf q}}
\left(b_{\lambda{\bf q}}^\dag b_{\lambda{\bf q}}+\frac1{2} \right), 
\end{equation}
and  ${\cal H}_p$ and  ${\cal H}_d$ defined by
Eqs.~(\ref{EqElPhP}) and (\ref{EqElPhD}), respectively.
The $t$-$J$ model ${\cal H}_{t\mbox{-}J}$
is defined on a square lattice in Sec.~\ref{KL-theory}  but it
should be defined on a quasi-two-dimensional lattice here.
Although no interlayer coupling is included in ${\cal H}_{t\mbox{-}J}$,
the nature of quasi-two-dimensional AF spin fluctuations
is phenomenologically considered,
which plays one of the most crucial roles in the softening of
Cu-O bond stretching modes, as is examined below. 

In the absence of ${\cal H}_p$ and  ${\cal H}_d$,
the Green function for phonons  is given by
\begin{equation}
D_\lambda^{(0)}(i\omega_l,{\bf q})=
\frac{2 \omega_{\lambda{\bf q}} }{
(i\omega_l)^2 - \omega_{\lambda{\bf q}}^2} .
\end{equation}
When the self-energy for phonons is denoted by
$\Sigma_{\lambda}^{({\rm ph})}(i\omega_l,{\bf q})$,
which can be perturbatively calculated 
in terms of ${\cal H}_p$ and  ${\cal H}_d$,
the renormalized Green function for phonons  is given by
\begin{eqnarray}\label{EqSoft}
D_\lambda(i\omega_l,{\bf q}) &=&
D_\lambda^{(0)}(i\omega_l,{\bf q}) 
\nonumber \\ &&  +
D_\lambda^{(0)}(i\omega_l,{\bf q})
\Sigma_{\lambda}^{({\rm ph})}(i\omega_l,{\bf q})
D_\lambda(i\omega_l,{\bf q})
\nonumber \\ &=&
\frac{2 \omega_{\lambda{\bf q}} }{
(i\omega_l)^2 - \omega_{\lambda{\bf q}}^2
- 2 \omega_{\lambda{\bf q}}
\Sigma_{\lambda}^{({\rm ph})}(i\omega_l,{\bf q})}. \qquad 
\end{eqnarray}
%
The renormalized energy of phonons, which is
denoted by $\omega_{\lambda{\bf q}}^*$,  is given by
\begin{equation}
(\omega_{\lambda{\bf q}}^*)^2 - \omega_{\lambda{\bf q}}^2
- 2 \omega_{\lambda{\bf q}}
\Sigma_{\lambda}^{({\rm ph})}(\omega_{\lambda{\bf q}}^*+i0,{\bf q}) =0 .
\end{equation}
%
The renormalization of phonon energies is given by 
%
\begin{eqnarray}
\Delta \omega_{\lambda{\bf q}} &=&
\omega_{\lambda{\bf q}}^* - \omega_{\lambda{\bf q}}
\nonumber \\ &= &
\Sigma_{\lambda}^{({\rm ph})}(\omega_{\lambda{\bf q}}+i0,{\bf q}) 
\nonumber \\ && \quad 
- \bigl[\Sigma_{\lambda}^{({\rm ph})}
(\omega_{\lambda{\bf q}}+i0,{\bf q}) \bigr]^2/
2\omega_{\lambda{\bf q}} 
+ \cdots , \qquad
\end{eqnarray}
unless $|\Sigma_{\lambda}^{({\rm ph})}
(\omega_{\lambda{\bf q}}+i0,{\bf q})|$
is larger than $|\omega_{\lambda{\bf q}}|$.

As well as AF spin fluctuations are developed around ${\bf Q}$
due to $I_s(i\omega_l,{\bf q})$, 
$d\gamma$-wave superconducting (SC) and 
charge bond order (CBO) fluctuations are also developed due to 
$I_s(i\omega_l,{\bf q})$ or $I_s^*(i\omega_l,{\bf q})$.
Although charge fluctuations are never much developed, 
charge-channel fluctuations  can also contribute to the
softening of phonons, as well as AF, SC, and CBO fluctuations,
provided that vertex corrections for the dual spin operator 
in spin, SC, and CBO channels are properly treated.
According to the previous paper, \cite{novel-el-ph}
the softening of the half-breathing modes
is mainly caused by the charge-channel fluctuations.
Since the charge-channel fluctuations are significant
in the metallic phase,
the softening is large in the metallic phase but is
small in the insulating phase.

Since phonons can couple with two lines or two channels of spin fluctuations 
to the lowest or first order in the dual-spin operator, 
as is shown in Eqs.~(\ref{EqElPhP}) and (\ref{EqElPhD}), 
AF spin fluctuations around $(\pm 3 \pi/4a, \pm\pi/a)$ and
$( \pm\pi/a, \pm 3 \pi/4a)$ in 2DBZ can play a significant role
in the softening of Cu-O bond stretching modes
around $(\pm\pi/2a, 0)$ and $(0, \pm\pi/2a)$ in 2DBZ.
According to Mermin and Wagner, \cite{mermin}
if  the N\'{e}el temperature $T_N$ were nonzero in two dimensions
integrated effects of two-dimensional critical AF spin fluctuations
would be divergent at $T_N$, which leads to a conclusion that
 $T_N$ must be zero in two dimensions.
Their argument implies that quasi-two-dimensional critical
AF spin fluctuations can play a crucial role in the softening, at least,
in an AF critical region of cuprate oxide superconductors provided that
the anisotropy of critical AF spin fluctuations is large.
In order to examine how crucial a role the anisotropy plays in the softening, 
it is more convenient to use
a phenomenological expression for the spin susceptibility,
which includes explicitly the anisotropy factor for AF spin
fluctuations, than to
calculate microscopically the spin susceptibility 
for quasi-two-dimensional systems.
The superexchange interaction
$J({\bf q})$, which is given by Eq.~(\ref{EqSuperJ-2}), 
has broad peaks at $(\pm\pi/a, \pm\pi/a)$ in 2DBZ and 
the exchange interaction
$J_Q(0,{\bf q})$, which is given by Eq.~(\ref{EqJQ}), 
has sharp peaks at nesting wave numbers
of the Fermi surface. Therefore, it is assumed in this paper that 
$I_s(0,{\bf q})$ is maximal at
${\bf Q}=\left(\pm 3\pi/4a, \pm\pi/a,Q_z\right)$ and
$\left(\pm\pi/a, \pm 3\pi/4a, Q_z\right)$ or that
$\chi_s(0,{\bf q})$  is maximal at ${\bf Q}$, and that
the spin susceptibility  (\ref{EqSus}) is approximately but well
described  by
\begin{equation}\label{EqSus1}
\chi_s\left(i\omega_l, {\bf Q} + {\bf q}\right)
= \frac{\chi_s(0,{\bf Q})\kappa^2}
{\displaystyle
\kappa^2 \!+\!(q_\parallel a)^2 \!+\! \delta^2 (q_z c)^2 \!+\! 
\frac{|\omega_l|}{\Gamma_{\rm AF} } },
\end{equation}
around each of  ${\bf Q}$'s, with
${\bf q}_\parallel =(q_x,q_y)$ being the component parallel to
CuO$_2$ planes, $q_z$ the component perpendicular to CuO$_2$ planes,
$c$ the lattice constant along the $z$ axis, and
$\Gamma_{\rm AF}$ the energy scale of AF spin fluctuations.
The anisotropy factor $\delta$ is introduced to consider
quasi-two-dimensional AF spin fluctuations.
The correlation length within the $x$-$y$ plane is
$a/\kappa$ and that  along the $z$ axis is $\delta c /\kappa$.
A cut-off  $q_c=\pi/3a$ is introduced in such a way that
$\chi_s\left(i\omega_l, {\bf Q} + {\bf q}\right)=0$
for $|q_x|>q_c$ or $|q_y|>q_c$.
The anisotropy of the lattice constants
plays no role when $\delta$
and $q_c$ are defined in these ways.

When AF spin fluctuations are only considered,
the self-energy for phonons 
is given by \cite{novel-el-ph}
\begin{eqnarray}\label{EqSoftOmega}
\Sigma_{\lambda}^{({\rm ph})}(i\omega_l,{\bf q}) &=&
- \frac{\hbar^2}{2 M_p \omega_{\lambda{\bf q}} } 
\frac{3}{4^2}  \! \sum_{\Gamma\Gamma^\prime}
Y_{\Gamma}({\bf q}) Y_{\Gamma^\prime}({\bf q}) 
\nonumber \\ && \qquad \times
X_{\Gamma\Gamma^\prime}(i\omega_l,\!{\bf q}),
\end{eqnarray}
with
%
%
\begin{equation}
Y_{\Gamma}({\bf q}) = \bar{\eta}_{s}({\bf q}) \hspace{-2pt}
\left[C_p v_{p,\lambda{\bf q}} 
\eta_{\Gamma} \left(\mbox{$\frac{1}{2}{\bf q}$}\right)   
\!+\!  C_d v_{d,\lambda{\bf q}}\sqrt{
\displaystyle \frac{M_p}{M_d} } \right], 
\end{equation}
and
\begin{eqnarray}
\label{EqX} &&\hspace*{-0.5cm}
X_{\Gamma\Gamma^\prime}(i\omega_l,{\bf q})
=
\frac{k_B T}{N} \sum_{\omega_{l^\prime}{\bf p}}
\eta_{\Gamma}({\bf p}) \eta_{\Gamma^\prime}({\bf p}) 
%
\chi_s \! \left(i\omega_{l^\prime}, 
{\bf p} \!+\! \mbox{$\frac{1}{2}$}{\bf q}\right)
\nonumber \\ && \hspace*{2.0cm} \times
\chi_s \left(-i\omega_{l^\prime}-i\omega_l, 
-{\bf p} \!+\! \mbox{$\frac{1}{2}$}{\bf q}\right) .
\end{eqnarray}
In Eq.~(\ref{EqX}), two $\chi_s$'s appear because of
the dual-spin operator.
It should be noted that
$2{\bf Q}$'s are equivalent to
$(\pm \pi/2a,0)$ and $(0, \pm \pi/2a)$:
%
$2{\bf Q} - {\bf G} = (\pm \pi/2a,0) \mbox{~and~} (0, \pm \pi/2a)$,
%
with ${\bf G} = (\pm 2\pi/a, 0)$ and  $(0,\pm 2\pi/a)$
being reciprocal lattice vectors in 2DBZ.
Then, Cu-O bond stretching modes
around $(\pm \pi/2a,0)$ and $(0,\pm \pi/2a)$ in 2DBZ
can be soft provided that AF fluctuations around 
${\bf Q}=\left(\pm 3\pi/4a, \pm\pi/a\right)$ and
$\left(\pm\pi/a, \pm 3\pi/4a \right)$ in 2DBZ are developed.

\begin{figure*}[t]
\centerline{
\includegraphics[width=7.0cm]{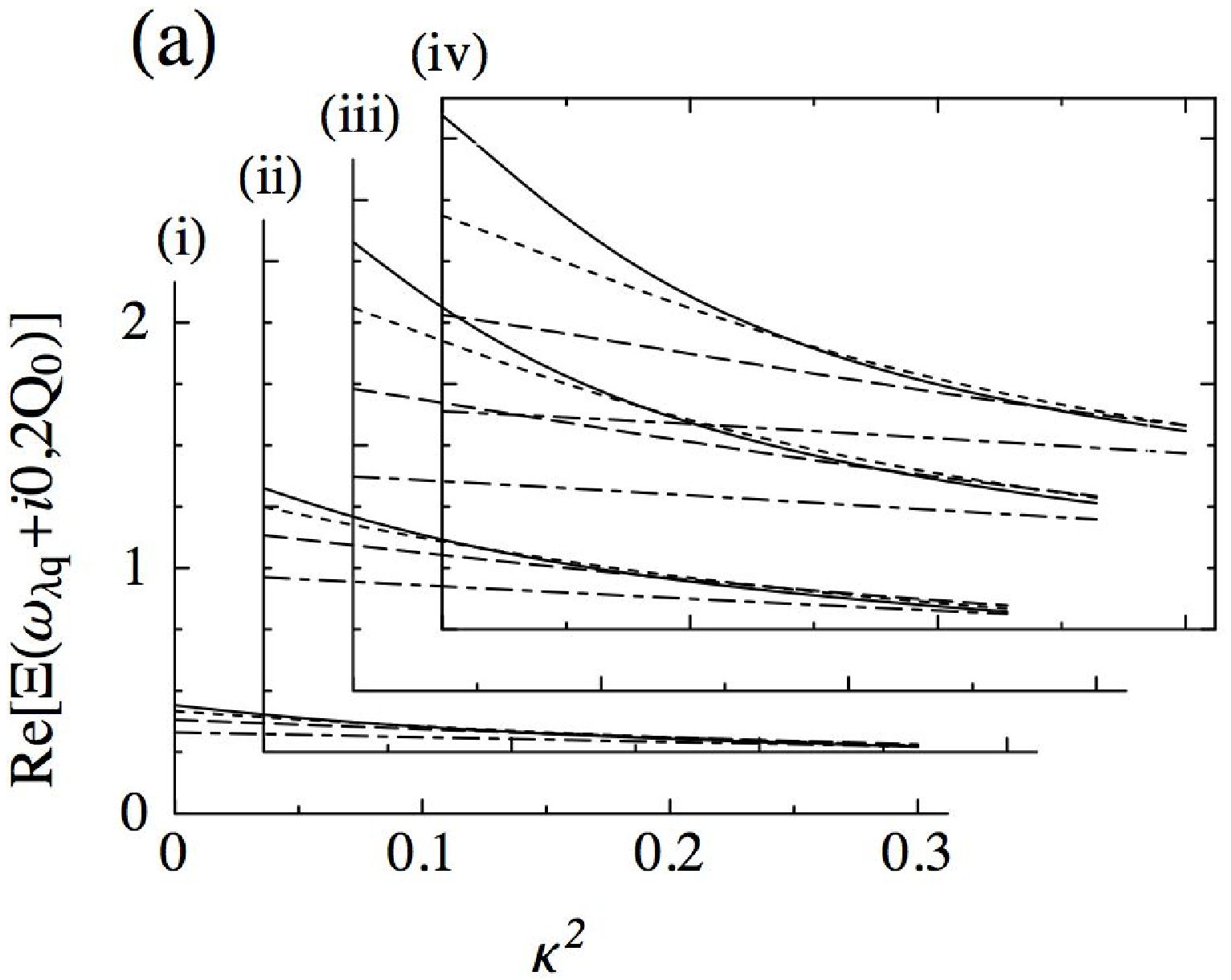}
\hspace{1cm}
\includegraphics[width=7.0cm]{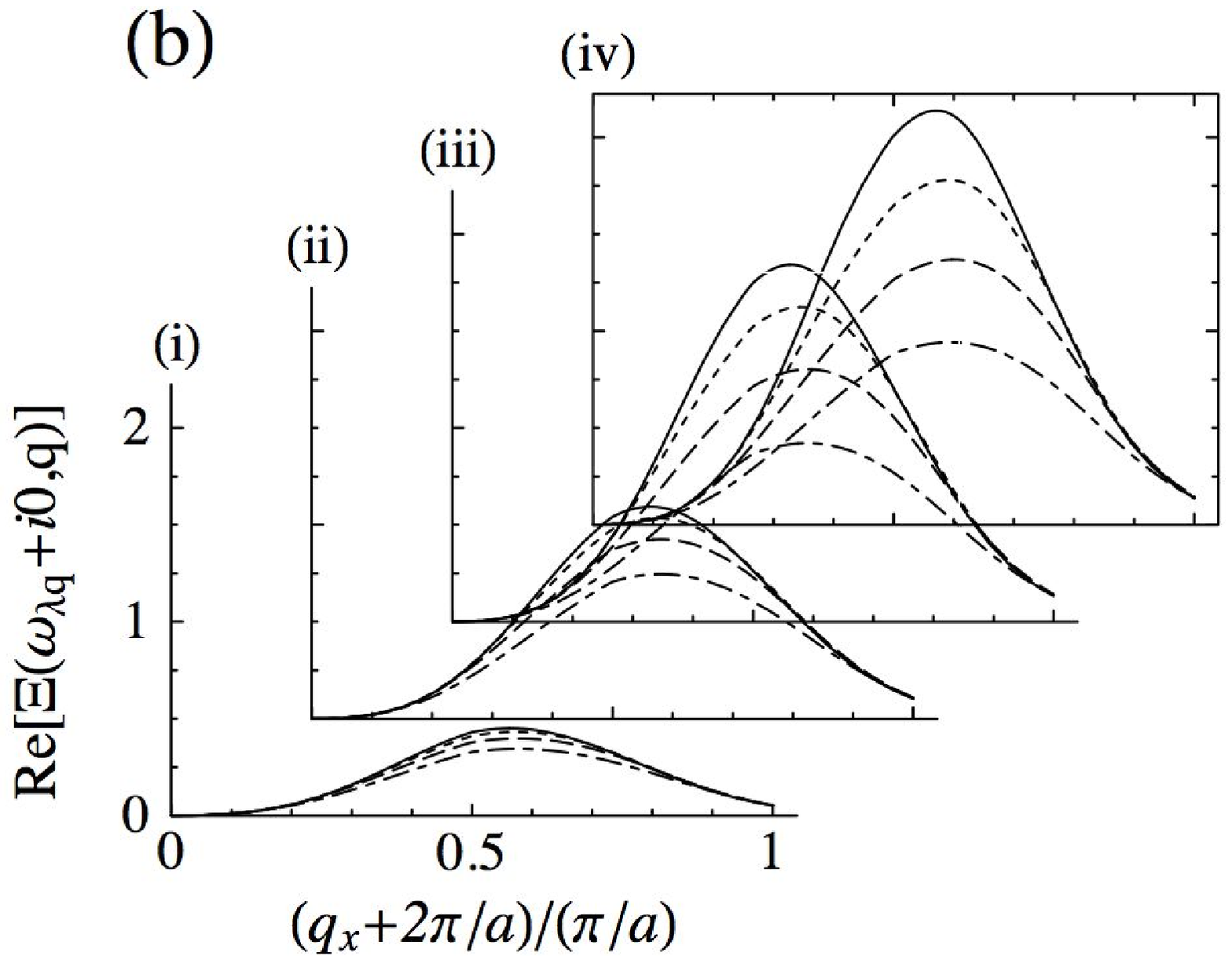}}
\caption[1]{
(a) $\mbox{Re}\bigl[\Xi(\omega_{\lambda{\bf q}}+i0,2{\bf Q}_0)\bigr]$ 
as a function of $\kappa^2$, with ${\bf Q}_0=(-3\pi/4a,\pi/a)$,
and (b) $\mbox{Re}\bigl[\Xi (\omega_{\lambda{\bf q}}+i0,{\bf q})\bigr]$ 
as a function of $q_x$ $(-2\pi/a \le q_x\le -\pi/a)$
for $q_y=2\pi/a$.  For the anisotropy factor, (i) $\delta=1$, 
(ii) $\delta=10^{-1/2}$, (iii) $\delta=10^{-1}$, and 
(iv) $\delta=10^{-3}$.
In each figure, solid, dotted, dashed, and dashed chain lines are for
$\omega_{\lambda{\bf q}}/\Gamma_{\rm AF}=$ 
0.2, 0.4, 0.8, and 1.6, respectively. 
}
\label{fig_Xi}
\end{figure*}

Since Cu-O bond stretching modes around $2{\bf Q}$ are considered, 
the vibrations of Cu ions are ignored, that is, it is assumed that
\begin{equation}
|C_d v_{d,\lambda{\bf q}}|\sqrt{M_p/M_d }=0,
\end{equation}
and
\begin{equation}
|C_p v_{p,\lambda{\bf q}} | = 
c_p \mbox{~eV}/\mbox{\AA},
\end{equation}
where $c_p$ is a dimensionless constant and it is likely 
\cite{novel-el-ph}
\begin{equation}
c_p=O(1).
\end{equation}
Since the contribution from small ${\bf p}$ is large in 
the summation over ${\bf p}$ in 
Eq.~(\ref{EqX}), only the contribution from 
the $\Gamma=s$ channel is considered.
Then, it follows that
\begin{equation}\label{EqXi1}
\Sigma_{\lambda}^{({\rm ph})}(\omega_{\lambda{\bf q}}+i0,{\bf q}) 
=- A_{\bf q} \Xi(\omega_{\lambda{\bf q}}+i0,{\bf q}) ,
\end{equation}
with
\begin{equation}\label{EqAq}
A_{\bf q} =
\frac{\hbar^2}{2 M_p \omega_{\lambda{\bf q}} } 
\frac{3}{4^2} \Gamma_{\rm AF} 
\left[\chi_s(0,{\bf Q})\kappa^2\right]^2
|C_p v_{p,\lambda{\bf q}} |^2 ,
\end{equation}
and 
\begin{equation}\label{EqXi2}
\Xi(\omega_{\lambda{\bf q}}+i0, {\bf q}) =
\bar{\eta}_s^2({\bf q})
\eta_{s}^2 \left(\mbox{$\frac{1}{2}{\bf q}$}\right)   
\frac{X_{ss}(\omega_{\lambda{\bf q}}+i0, {\bf q})}
{\Gamma _{\rm AF} \left[\chi_s({\bf Q})\kappa^2\right]^2}.
\end{equation}
%
%
It should be noted that $\Xi(i\omega_l, {\bf q})$ is defined 
as a dimensionless quantity. 
The effective transfer integral between nearest neighbors for
the Gutzwiller quasiparticles is 
\begin{equation}
t^* = t/\tilde{\phi}_\gamma.
\end{equation}
According to Eq.~(\ref{EqGamma2}), 
a plausible number for $t^*$ is 
\begin{equation}
|t^*| \simeq W^*/8\simeq 40\mbox{-50 meV}.
\end{equation}
According to a microscopic calculation for the spin susceptibility, 
it follows that
$\Gamma_{\rm AF}/|t^*|= O(1)$
and 
$\chi_{s}(0,{\bf Q}) \kappa^2 |t^*| = O(1)$.
%
%
It is assumed, for the sake of simplicity, that
the energy of Cu-O bond stretching modes is constant and is as large as
\begin{equation}
\omega_{\lambda{\bf q}} = 50\mbox{ meV} .
\end{equation}
Then, 
$A_{\bf q}$ defined by Eq.~(\ref{EqAq}) is approximately given by
\begin{eqnarray}\label{EqSizeA}
A_{\bf q}  &\simeq&
10 \times c_p^2 \frac{\Gamma_{\rm AF}}{|t^*|} 
\left[\chi_s(0,{\bf Q}) \kappa^2 |t^*|
\right]^2 \mbox{~meV} 
\nonumber \\ &\simeq& 
10 c_p^2 \mbox{~meV}. 
\end{eqnarray}
In this paper, 
$T=0$~K is assumed in the $\omega_{l^\prime}$ sum of Eq.~(\ref{EqX}).
 
The softening around one of $2{\bf Q}$'s
or $2{\bf Q}_0$, with
${\bf Q}_0=(-3\pi/4a,\pi/a)$ in 2DBZ 
is considered;
$2{\bf Q}_0$ is equivalent to $(\pi/2a,0)$.
Figure~\ref{fig_Xi} shows 
the dependence of $\Xi(\omega_{\lambda{\bf q}}+i0,{\bf q})$ on
$\kappa$, $\delta$, $\Gamma_{\rm AF}$, and ${\bf q}$;
Fig.~\ref{fig_Xi}(a) and Fig.~\ref{fig_Xi}(b) show
$\mbox{Re}[\Xi(\omega_{\lambda{\bf q}}+i0, 2{\bf Q}_0)]$
as a function of $\kappa^2$ and
$\mbox{Re}[\Xi(\omega_{\lambda{\bf q}}+i0, {\bf q})]$
as a function of $q_x$, respectively, 
for several sets of $\delta$ and 
$\omega_{\lambda{\bf q}}/\Gamma_{\rm AF}$.
According to Fig.~\ref{fig_Xi}(b),
$\Xi(\omega_{\lambda{\bf q}}+i0, {\bf q}) $ has a maximum, that is, 
$\mbox{Re}\bigl[\Sigma_{\lambda}^{({\rm ph})}
(\omega_{\lambda{\bf q}}+i0,{\bf q}) \bigr]$
has a minimum around $2{\bf Q}_0$ as a 
function of ${\bf q}$.  
According to Eq.~(\ref{EqSizeA}), 
Fig.~\ref{fig_Xi}(a), and Fig.~\ref{fig_Xi}(b),
it is likely that the softening at $2{\bf Q}_0$
is as large as $-$(10-20) meV for
$\kappa^2 \ll 1$ and $\delta \ll 1$.
It should be noted that the softening can only be large 
provided that $\kappa^2 \ll 1$ and $\delta \ll 1$, as is implied by
Mermin and Wagner's argument; \cite{mermin}
integrated effects on 
the softening are never divergent even in the limit of 
$\kappa\rightarrow 0$ and $\delta\rightarrow0$.

It is definite that
$\kappa^2 \ll1$ in the critical region of antiferromagnetic ordering or
spin density wave (SDW), and it is certain
that the anisotropy is as large as $\delta <10^{-3}$ in cuprate oxide
superconductors.
Then, the second-harmonic effect of AF spin fluctuations
can explain the observed softening \cite{pintschovius,reznik}
as large as $-$(10-20)~meV around $2{\bf Q}_0$ or $2{\bf Q}$.

Since the softening is small when $\kappa^2$ is large or
AF spin fluctuations are not developed, 
it must be small in 
over-doped cuprate oxide superconductors,
whose doping concentrations
are larger than those of optimal-doped ones.
When AF spin fluctuations are developed similarly or differently between 
$\left(\pm 3\pi/4a, \pm\pi/a\right)$  and  
$\left(\pm\pi/a, \pm 3\pi/4a \right)$ 
because of the anisotropy of the Fermi surface within 2DBZ,
the softening must also occurs similarly or differently between 
the $x$ and $y$ axes or between
$\left(\pm\pi/2a,0\right)$ and $\left(0,\pm\pi/2a\right)$.
These two predictions are consistent with observations.
\cite{pintschovius,reznik}

\section{Stripes and checker boards}  
\label{stripes}

Since the $4a$-period and $4a\!\times\! 4a$-period correspond to
 $2{\bf Q}$, with 
 ${\bf Q}=\left(\pm 3\pi/4a, \pm\pi/a\right)$ and
$\left(\pm\pi/a, \pm 3\pi/4a \right)$,
a plausible scenario for the stripes and checker boards
is that the complete softening is followed by the stabilization of 
CDW with $2{\bf Q}$. 
In general, the $2{\bf Q}$ component of 
the density of states, $\rho_{2{\bf Q}}(\varepsilon)$,  as a function of
$\varepsilon$ is composed of
symmetric and asymmetric components with respect to 
the chemical potential or $\varepsilon=0$.
The asymmetric component is large
when CDW with $2{\bf Q}$ is  stabilized 
as a fundamental $2{\bf Q}$ effect.
According to an experiment, \cite{howald}
the symmetric component is larger than the asymmetric one, which
contradicts the scenario of CDW even if 
the softening of the $2{\bf Q}$ modes is large and the $2{\bf Q}$
fluctuations are well developed.
On the other hand, the symmetric component is large 
when the $2{\bf Q}$ modulation  is due to a
simple second-harmonic effect
of an ordered SDW with ${\bf Q}$. \cite{FJO-FFLO} 
The second-harmonic effect of the SDW can explain
the observed almost symmetric 
$\rho_{2{\bf Q}}(\varepsilon)$.
When stripes and checker-boards are really static orders, 
stripes must be due to single-${\bf Q}$ SDW and checker-boards
must be due to double-${\bf Q}$ SDW.
It is predicted that
magnetizations of the two waves must be orthogonal 
to each other in double-${\bf Q}$ SDW. \cite{ortho1,ortho2}
It is interesting to examine whether 
the prediction actually holds in cuprate oxide superconductors.

The appearance or stabilization of SDW is a transition rather than a crossover.
However, no specific heat anomaly has been reported so far
except for the anomaly due to superconductivity.
The absence of any specific heat anomaly implies that,
even if SDW is stabilized,
SDW is never a homogeneous phase but is an inhomogeneous phase,
which is composed of many domains.
If the transition temperatures of SDW can be different 
in different domains, no significant specific anomaly can be observed.
It is plausible that SDW is 
an disorder-induced SDW. \cite{FJO-disorder}

On the other hand, it is proposed  \cite{kiverson}
that a stripe or a checker-board at rather high temperatures 
must be an exotic ordered state, that is,
a fluctuating state in a quantum disordered phase.
It should be examined 
whether it is actually such an exotic state.
Another possibility is that it is 
a rather normal low-energy fluctuating state, whose
energy scale is as small as that of the soft phonons.
The other possibility is the disorder-induced SDW,
\cite{FJO-disorder} 
which can behave as a fluctuating state because it is inhomogeneous.

\section{Attractive interaction} 
\label{SecDisuss}

Although the electron-phonon interaction plays 
no or only a minor role 
in the formation of $d\gamma$-wave Cooper pairs,
as is discussed in the Introduction, 
isotope shifts of $T_c$ can arise from the depression of
superconductivity by the $2{\bf Q}$ fluctuations, 
whose development depends on the mass of O ions.

In Kondo-lattice theory, 
cuprate oxide superconductors can be relevantly treated as
one of the typical Kondo lattices.
According to Eq.~(\ref{EqSF}), which is one of the most crucial results
of Kondo-lattice theory, two mechanisms of  attractive interactions,
the  spin-fluctuation mechanism and  the exchange-interaction mechanism,
are essentially the same as each other.
However, the attractive interaction mediated by low-energy spin fluctuations
such as those observed by low-energy neutron inelastic scatterings
or those described by the phenomenological spin susceptibility
(\ref{EqSus}) is physically different from the attractive interaction due
to the superexchange interaction, which arises from the exchange
of a pair excitation of electrons in spin channels across the Hubbard gap
because the energy scales of spin fluctuations or spin excitations are
totally different from each other in the two physical processes.
The main part of the attractive interaction in cuprate oxide superconductors
must be the superexchange interaction rather than the interaction
mediated by low-energy AF spin fluctuations.
Since the superexchange interaction
 is as strong as $J=-(0.10$-0.15)~eV, \cite{SuperJ}
observed high $T_c$ can be easily reproduced, as is discussed
in the Introduction. 

~~

\section{Conclusion}
\label{SecConclusion}
In cuprate oxides superconductors,
the  electron-phonon interaction arising from the modulation of 
the superexchange interaction by lattice vibrations is strong enough 
to cause the softening of not only the half-breathing modes
around $(\pm\pi/a, 0)$ and $(0,\pm\pi/a)$ 
in the two-dimensional Brillouin zone
but also Cu-O bond stretching modes 
around $(\pm\pi/2a, 0)$ and $(0,\pm\pi/2a)$.
Although the softening of the bond stretching modes is responsible
for stripe and checker-board fluctuations in charge channels, 
the stabilization of a charge density wave state following
the complete softening of the bond stretching modes can never 
be any relevant scenario for ordered stripe and checker-board states.
The ordered stripe or checker-board state must be simply
a single-{\bf Q} or double-{\bf Q} spin density wave state, whose
${\bf Q}$'s are $\left(\pm 3\pi/4a, \pm\pi/a\right)$ and
$\left(\pm\pi/a, \pm 3\pi/4a \right)$.
The  strong electron-phonon interaction can play no or only a minor role
in the binding of $d\gamma$-wave Cooper pairs
in cuprate oxide superconductors, because the attractive interaction
arising from the virtual exchange of a phonon is never strong
between quasi-particles on nearest-neighbor Cu ions on a CuO$_2$ plane.
However, isotope shifts of $T_c$ can arise from the depression of
superconductivity by the stripe or checker-board fluctuations.
Since the superexchange interaction is as strong as 
$J=-(0.10\mbox{-}0.15)$~eV between nearest-neighbor Cu ions, 
the superexchange interaction must be mainly responsible for the binding of 
the $d\gamma$-wave Cooper pairs in cuprate oxide superconductors.


\begin{thebibliography}{}
\bibitem{bednorz}
J. G. Bednortz and K. A. M\"{u}ller, 
Z. Phys. B {\bf 64}, 189 (1986).
%
\bibitem{McQ1} 
R. J. McQueeney,  
Y. Petrov, T. Egami, M. Yethiraj, G. Shirane, and Y. Endoh,
Phys. Rev. Lett. {\bf 82}, 628 (1999).
%
\bibitem{Pint1}
L. Pintschovius and M. Braden,
Phys. Rev. B {\bf 60}, R15039 (1999).
%
\bibitem{McQ2}
R. J. McQueeney, 
J. L. Sarrao, P. G. Pagliuso, P. W. Stephens, and R. Osborn,  
Phys. Rev.  Lett. {\bf 87}, 077001 (2001).
%
\bibitem{Pint2}
L. Pintschovius, 
W. Reichardt, M. Braden, G. Dhalenne, and A. Revcolevschi, 
Phys. Rev. B {\bf 64}, 094510 (2001).
%
\bibitem{Braden} 
M. Braden, 
W. Reichardt, S. Shiryaev and S. N. Barilo,
Physica C {\bf 378}-{\bf 381}, 89 (2002).
%
\bibitem{pintschovius}
L. Pintschovius, 
D. Reznik, W. Reichardt, Y. Endoh, H. Hiraka, J. M. Tranquada, H. Uchiyama, T. Masui, and S. Tajima,
Phys. Rev. B {\bf 69}, 214506 (2004).
%
\bibitem{reznik} 
D. Reznik, 
L. Pintschovious, M. Ito, S. Iikubo, M. Sato, H. Goka, M. Fujita, K. Yamada, G. D. Gu, and J. M. Tranquada,
Nature, {\bf 440}, 1170 (2006). 
%
\bibitem{johnson}
P. D. Johnson, 
 T. Valla, A. V. Fedorov, Z. Yusof, B. O. Wells, Q. Li, A. R. Moodenbaugh,
 G. D. Gu, N. Koshizuka, C. Kendziora, Sha Jian, and D. G. Hinks,
Phys. Rev. Lett {\bf 87}, 177007 (2001).
%
\bibitem{tsato}
T. Sato, 
 H. Matsui, T. Takahashi, H. Ding, H.-B. Yang, S.-C. Wang, T. Fujii,
 T. Watanabe, A. Matsuda, T. Terashima, and K. Kadowaki, 
Phys. Rev. Lett. {\bf 91}, 157003 (2003).
%
\bibitem{isotope}
J. P. Franck, 
{\it Physical Properties of High Temperature Superconductors IV},
 ed. D.M. Ginsberg (World Scientific, Singapore, 1994)  p189. 
%
%
\bibitem{Hubbard1} 
J. Hubbard, 
Proc. Roy. Soc. London Ser. A\ {\bf 276}, 238 (1963).
%
\bibitem{Hubbard2} 
J. Hubbard,
Proc. Roy. Soc. London Ser. A {\bf 281}, 401 (1964).
%
\bibitem{Gutzwiller1} 
M. C. Gutzwiller, 
Phys. Rev. Lett. {\bf 10}, 159 (1963).
%
\bibitem{Gutzwiller2} 
M. C. Gutzwiller, 
Phys. Rev. {\bf 134}, A923 (1963).
%
\bibitem{Gutzwiller3} 
M. C. Gutzwiller, 
Phys. Rev. {\bf 137} A1726, (1965).
%
\bibitem{OhkawaSlave}
F. J. Ohkawa:
J. Phys. Soc. Jpn. {\bf 58}, 4156 (1989).
%
\bibitem{Mapping-1} 
F. J. Ohkawa, 
Phys. Rev. B {\bf 44}, 6812 (1991).
%
\bibitem{Mapping-2} 
F. J. Ohkawa, 
J. Phys. Soc. Jpn. {\bf 60}, 3218 (1991).
%
\bibitem{Mapping-3} 
F. J. Ohkawa, 
J. Phys. Soc. Jpn. {\bf 61}, 1615 (1992).
%
\bibitem{yosida}
K. Yosida,
Phys. Rev. {\bf 147}, 223 (1966).
%
\bibitem{wilsonKG}
K. G. Wilson, 
Rev. Mod. Phys. {\bf 47}, 773 (1975).
%
\bibitem{ohkawaM-I}
F. J. Ohkawa, cond-mat/0606644.
It is possible that under the SSA
the normal Fermi-liquid ground state is degenerate
with a non-normal Fermi liquid state provided that the chemical potential
is just at the singular point of the density of states, for example,
just at the logarithmic van Hove singularity 
of two dimensional systems.
%
\bibitem{georges} 
A. Georges and G. Kotliar,
Phys. Rev. B {\bf 45}, 6479 (1992).
%
\bibitem{kakehashi}
Y. Kakehashi and P. Fulde,
Phys. Rev. B {\bf 69}, 45101 (2004).
%
\bibitem{Metzner}
W. Metzner and D. Vollhardt,
Phys. Rev. Lett. {\bf 62}, 324 (1989).
%
\bibitem{gamma1} 
J. W. Loram,  
K. A. Mirza, J. R. Cooper, and W. Y. Liang,
Phys. Rev. Lett. {\bf 71}, 1740 (1993).
%
\bibitem{gamma2} 
N. Momono,  
M. Ido, T. Nakano, M. Oda, Y. Okajima, and K. Yamaya,
Physica C {\bf 233}, 395 (1994).
%
\bibitem{Luttinger1}
J. M. Luttinger and J. C. Ward:
Phys. Rev. {\bf 118}, 1417 (1960);
%
\bibitem{Luttinger2}
J. M. Luttinger,
Phys. Rev. {\bf 119}, 1153 (1960). 
%
\bibitem{OhSupJ1} 
F. J. Ohkawa, Phys. Rev. B {\bf 59}, 8930 (1999). 
%
\bibitem{hirsch}
J. E. Hirsch,
Phys. Rev. Lett. {\bf 54}, 1317 (1985).
%
\bibitem{SuperJ}
K. B. Lyons,  
P. A. Fleury, L. F. Schneemeyer, and J. V. Waszczak,
Phys. Rev. Lett. {\bf 60}, 732 (1988).
%
\bibitem{highTc1} 
F. J. Ohkawa, 
Jpn. J. Appl. Phys. {\bf 26}, L652 (1987).
%
\bibitem{highTc2} 
F. J. Ohkawa, 
J. Phys. Soc. Jpn. {\bf 56}, 2267 (1987).
%
\bibitem{novel-el-ph}
F. J. Ohkawa,
Phys. Rev. B {\bf 70}, 184514 (2004).
%
\bibitem{howald}
C. Howald,  
H. Eisaki, N. Kaneko, M. Greven, and A. Kapitulnik,
Phys. Rev. B {\bf 67}, 014533 (2003).
%
\bibitem{Vershinin} 
M. Vershinin, 
S. Mirsa, S. Ono, Y. Abe, Y. Ando, and A. Yazdani,
Science {\bf 303}, 1995 (2004).
%
\bibitem{Hanaguri}
T. Hanaguri, 
G. Lupien, Y. Kohsaka, D.-H. Lee, M. Azumas, M. Takano, H. Takagi, and J. C. Davis,
Nature {\bf 430}, 1001 (2004).
%
\bibitem{Momono}
N. Momono, 
A. Hashimoto,Y. Kobatake, M. Oda, and M. Ido, 
J. Phys. Soc. Jpn. {\bf 74}, 2400 (2005).
%
\bibitem{McElroy}
K. McElroy, 
D.-H. Lee, J. E. Hoffman, K. M. Lang, J. Lee, E. W. Hudson, H. Eisaki, S. Uchida, and J. C. Davis,
Phys. Rev. Lett. {\bf 94}, 197005 (2005).
%
\bibitem{kiverson}
S. A. Kiverson, 
I. P. Bindloss, E. Fradkin, V. Oganesyan, J. M. Tranquanda, A. Kapitulnik, and C. Howald,
Rev. Mod. Phys. {\bf 75}, 1202 (2003).
%
\bibitem{band1}
T. Takegahara, H. Harima, and Y. Yanase,
Jpn. J. Appl. Phys. Part 1, {\bf 26}, L352 (1987)
%
\bibitem{band2}
N. Hamada,  
S. Massidda, A. J. Freeman, and J. Redinger,
Phys. Rev. B {\bf 40}, 4442 (1989).
%
\bibitem{band3}
O. K. Andersen,  
O. Jepsen, A. I. Liechtenstein, and I. I. Mazin,
Phys. Rev. B {\bf 49}, 4145 (1994).
%
\bibitem{ZhangRice}
F. C. Zhang and T. M. Rice,
Phys. Rev. B {\bf 37}, R3759 (1988). 
%
\bibitem{FJO-disorder}
F. J. Ohkawa,
J. Phys. Soc. Jpn. {\bf 74}, 3340 (2005).
%
\bibitem{ward}
J. C. Ward, 
Phys. Rev. {\bf 68}, 182 (1950).
%
\bibitem{moriya}
See, for example, T. Moriya, 
{\it Spin Fluctuations in Itinerant Electron Magnetism},
Springer Series in Solid-State Sciences, Vol. 56 (Springer-Verlag, 
Berlin, Heidelberg, New York, Tokyo, 1985).
%
\bibitem{mermin}
N. D. Mermin and H. Wagner,
Phys. Rev. Lett. {\bf 17}, 1133 (1966).
%
\bibitem{FJO-FFLO}
F. J. Ohkawa,
Phys. Rev. B {\bf 73}, 092506 (2006).
%
\bibitem{ortho1}
F. J. Ohkawa,
J. Phys. Soc. Jpn. {\bf 67}, 535 (1998).
%
\bibitem{ortho2}
F. J. Ohkawa,
Phys. Rev. B {\bf 66}, 014408 (2002).
%
\end{thebibliography}
\end{document}